\documentclass[12pt]{article}

\usepackage{epsf}
\usepackage{rotate}

\def\spose#1{\hbox to 0pt{#1\hss}}
\def\lsim{\mathrel{\spose{\lower 3pt\hbox{$\mathchar"218$}}
 \raise 2.0pt\hbox{$\mathchar"13C$}}}
\def\gsim{\mathrel{\spose{\lower 3pt\hbox{$\mathchar"218$}}
 \raise 2.0pt\hbox{$\mathchar"13E$}}}

\begin{document}

\begin{titlepage}

\begin{flushright}
CERN-TH/99-92\\
hep-ph/9903540
\end{flushright}

\vspace{1.5cm}
\begin{center}
\boldmath
\large\bf
Extracting CKM Phases from Angular Distributions of\\ 
\vspace{0.3truecm}
$B_{d,s}$ Decays into Admixtures of CP Eigenstates
\unboldmath
\end{center}

\vspace{1.2cm}
\begin{center}
Robert Fleischer\\[0.1cm]
{\sl Theory Division, CERN, CH-1211 Geneva 23, Switzerland}
\end{center}

\vspace{1.7cm}
\begin{abstract}
\vspace{0.2cm}\noindent
The time-dependent angular distributions of certain $B_{d,s}$ decays 
into final states that are admixtures of CP-even and CP-odd configurations 
provide valuable information about CKM phases and hadronic parameters. We
present the general formalism to accomplish this task, taking also into 
account penguin contributions, and illustrate it by considering a few 
specific decay modes. We give particular emphasis to the decay 
$B_d\to J/\psi\,\rho^0$, which can be combined with $B_s\to J/\psi\,\phi$ 
to extract the $B^0_d$--$\overline{B^0_d}$ mixing phase and -- if penguin 
effects in the former mode should be sizeable -- also the angle $\gamma$ 
of the unitarity triangle. As an interesting by-product, this strategy 
allows us to take into account also the penguin effects in the extraction 
of the $B^0_s$--$\overline{B^0_s}$ mixing phase from $B_s\to J/\psi\,\phi$.
Moreover, a discrete ambiguity in the extraction of the CKM angle $\beta$ 
can be resolved, and valuable insights into $SU(3)$-breaking effects can 
be obtained. Other interesting applications of the general formalism 
presented in this paper, involving $B_d\to\rho \rho$ and 
$B_{s,d}\to K^{\ast}\overline{K^\ast}$ decays, are also briefly noted.
\end{abstract}

\vfill
\noindent
CERN-TH/99-92\\
April 1999

\end{titlepage}

\thispagestyle{empty}
\vbox{}
\newpage
 
\setcounter{page}{1}

\section{Introduction}\label{intro}
Studies of CP violation in the $B$-meson system and the determination 
of the three angles $\alpha$, $\beta$ and $\gamma$ of the usual 
non-squashed unitarity triangle \cite{ut} of the Cabibbo--Kobayashi--Maskawa 
matrix (CKM matrix) \cite{ckm} are among the central targets of future 
$B$-physics experiments. During the recent years, several strategies
were proposed to accomplish this task \cite{revs}. In this context, also 
quasi-two-body modes $B_q\to X_1\,X_2$ of neutral $B_q$-mesons 
($q\in\{d,s\}$), where both $X_1$ and $X_2$ carry spin and continue to 
decay through CP-conserving interactions, are of particular interest 
\cite{dqstl,fd}. In this case, the time-dependent angular distribution 
of the decay products of $X_1$ and $X_2$ provides valuable information. 
For an initially, i.e.\ at time $t=0$, present $B_q^0$-meson, it can be 
written as 
\begin{equation}\label{ang}
f(\Theta,\Phi,\Psi;t)=\sum_k{\cal O}^{(k)}(t)g^{(k)}(\Theta,\Phi,\Psi),
\end{equation}
where we have denoted the angles describing the kinematics of the decay
products of $X_1$ and $X_2$ generically by $\Theta$, $\Phi$ and
$\Psi$. Note that we have to deal, in general, with an arbitrary number of 
such angles. The observables ${\cal O}^{(k)}(t)$ describing the
time evolution of the angular distribution (\ref{ang}) can be expressed 
in terms of real or imaginary parts of certain bilinear combinations of 
decay amplitudes. In the applications discussed in this paper, we will 
focus on $B_q\to [X_1\,X_2]_f$ decays, where $X_1$ and $X_2$ are 
both vector mesons, and $f$ denotes a final-state configuration with
CP eigenvalue $\eta_f$. It is convenient to analyse such modes  
in terms of the linear polarization amplitudes $A_0(t)$, $A_\parallel(t)$ 
and $A_\perp(t)$ \cite{pol}. Whereas $A_\perp(t)$ describes a CP-odd 
final-state configuration, both $A_0(t)$ and $A_\parallel(t)$ correspond 
to CP-even final-state configurations, i.e.\ to the CP eigenvalues $-1$ 
and $+1$, respectively. The observables of the corresponding angular 
distribution are given by
\begin{equation}\label{obs1}
\left|A_f(t)\right|^2\quad\mbox{with}\quad f\in\{0,\parallel,\perp\},
\end{equation}
as well as by the interference terms
\begin{equation}\label{obs2}
\Re\{A_0^\ast(t)A_\parallel(t)\}\quad\mbox{and}\quad
\Im\{A_f^\ast(t)A_\perp(t)\} \quad\mbox{with}\quad f\in\{0,\parallel\}.
\end{equation}
This formalism is discussed in more detail in \cite{ddf1}, 
where several explicit angular distributions can be found and appropriate 
weighting functions to extract their observables in an efficient way from 
the experimental data are given.

In the following considerations, the main role is played by neutral
$B_q\to [X_1\,X_2]_f$ decays, where the ``unevolved'' decay amplitudes 
can be expressed as
\begin{eqnarray}
A_f&=&{\cal N}_f\left[1-b_f\, e^{i\rho_f} e^{+i\omega}\right]\label{ampl}\\
\overline{A}_f&=&\eta_f\,{\cal N}_f\left[1-b_f\, e^{i\rho_f} 
e^{-i\omega}\right],\label{ampl-CP}
\end{eqnarray}
where $\omega$ denotes a CP-violating weak phase and ${\cal N}_f\equiv
|{\cal N}_f|e^{i\delta_f}$. Both $\rho_f$ and $\delta_f$ are CP-conserving
strong phases. In this case, the observables (\ref{obs1}) and (\ref{obs2})
allow us to probe the $B^0_q$--$\overline{B^0_q}$ mixing phase $\phi_q$ 
and the weak phase $\omega$, as we will show in this paper. Concerning 
practical applications, $\omega$ is given by one of the angles 
of the unitarity triangle. However, the observables specified in (\ref{obs1}) 
and (\ref{obs2}) are not independent from one another and do not provide 
sufficient information to extract $\phi_q$ and $\omega$, as well as the 
corresponding hadronic parameters, simultaneously. To this end, we have
to use an additional input. 

Usually, the weak phase $\omega$ is of central interest. If we fix the 
mixing phase $\phi_q$ separately, it is possible to determine $\omega$ -- and 
interesting hadronic quantities -- as a function of a {\it single} hadronic 
parameter in a {\it theoretically clean} way. If we determine this quantity, 
for instance, by comparing $B_q\to X_1\,X_2$ with an $SU(3)$-related mode, 
all remaining parameters, including $\omega$, can be extracted. If we are 
willing to make more extensive use of flavour-symmetry arguments, it is in 
principle possible to determine the $B^0_q$--$\overline{B^0_q}$ mixing 
phase $\phi_q$ as well. An example for such a strategy is given by the 
decay $B_d\to J/\psi\,\rho^0$, which can be combined with 
$B_s\to J/\psi\,\phi$ to extract the $B^0_d$--$\overline{B^0_d}$ mixing 
phase $\phi_d=2\beta$ and -- if penguin effects in the former mode should 
be sizeable -- also the angle $\gamma$ of the unitarity triangle. As an 
interesting by-product, this strategy allows us to take into account also 
the penguin effects in the extraction of the $B^0_s$--$\overline{B^0_s}$ 
mixing phase from $B_s\to J/\psi\,\phi$, which is an important issue for
``second-generation'' $B$-physics experiments at hadron machines. Moreover,
we may resolve a discrete ambiguity in the extraction of the CKM angle 
$\beta$, and may obtain valuable insights into $SU(3)$-breaking effects. 
Other interesting applications of the general formalism presented in this 
paper, involving $B_d\to\rho \rho$ and $B_{s,d}\to K^{\ast}
\overline{K^{\ast}}$ decays, are also briefly noted.

As the extraction of $\omega$ with the help of these modes involves 
``penguin'', i.e.\ flavour-changing neutral-current (FCNC) processes 
and relies moreover on the unitarity of the CKM matrix, it may well be 
affected by new physics. In such a case, discrepancies would show up 
with other strategies to determine this phase, for example with the 
theoretically clean extractions of $\gamma$ making use of pure ``tree'' 
decays such as $B_s\to D_s^\pm K^\mp$. Since no FCNC processes contribute 
to the decay amplitudes of these modes, it is quite unlikely that they -- 
and the extracted value of $\gamma$ -- are significantly affected by new 
physics.

The outline of this paper is as follows: in Section~\ref{EVOL}, the 
time-dependent observables of the $B_q\to X_1\,X_2$ angular distribution
are given. The strategies to extract CKM phases, as well as 
interesting hadronic parameters, with the help of these observables 
are discussed in Section~\ref{STRAT}. In Section~\ref{BPSI}, we focus
on the extraction of $\beta$ and $\gamma$ from $B_d\to J/\psi\,\rho^0$ 
and $B_s\to J/\psi\,\phi$. Further applications of the formalism developed 
in Sections~\ref{EVOL} and \ref{STRAT} are discussed in Section~\ref{APPL}, 
and the conclusions are summarized in Section~\ref{CONCL}.

\boldmath
\section{The Time Evolution of the Angular Distributions}\label{EVOL}
\unboldmath
In this section, we consider the general case of a neutral quasi-two-body 
decay $B_q\to[X_1\,X_2]_f$ into a final-state configuration
$f$ with CP eigenvalue $\eta_f$ that exhibits ``unmixed'' decay amplitudes
of the same structure as those given in (\ref{ampl}) and (\ref{ampl-CP}). 
If we use linear polarization states to characterize the final-state 
configurations as, for example, in \cite{ddf1}, we have 
$f\in\{0,\parallel,\perp\}$. 

At this point a comment on the angular distribution of the CP-conjugate 
decay $\overline{B_q^0}\to X_1\,X_2$, which is given by
\begin{equation}\label{ang-CP}
\overline{f}(\Theta,\Phi,\Psi;t)=\sum_k\overline{{\cal O}}^{(k)}(t)
g^{(k)}(\Theta,\Phi,\Psi),
\end{equation}
is in order. Since the meson content of the $X_1\,X_2$ states is the 
same whether these result from the $B_q^0$ or $\overline{B_q^0}$ decays, 
we may use the same generic angles $\Theta$, $\Phi$ and $\Psi$ to describe 
the angular distribution of their decay products. Within this formalism, 
the effects of CP transformations relating $B_q^0\to[X_1\,X_2]_f$ to 
$\overline{B_q^0}\to[X_1\,X_2]_f$ are taken into account by the CP 
eigenvalue $\eta_f$ appearing in (\ref{ampl-CP}), and do not affect the 
form of $g^{(k)}(\Theta,\Phi,\Psi)$. Therefore the same functions 
$g^{(k)}(\Theta,\Phi,\Psi)$ are present in (\ref{ang}) and (\ref{ang-CP}) 
(see also \cite{fd}). 

In view of applications to $B_s$ decays, we allow for a non-vanishing 
width difference $\Delta\Gamma_q\equiv\Gamma_{\rm H}^{(q)}-
\Gamma_{\rm L}^{(q)}$ between the $B_q$ mass eigenstates $B^{\rm H}_q$ 
(``heavy'') and $B^{\rm L}_q$ (``light''). In contrast to the $B_d$ case, 
this width difference may be sizeable in the $B_s$ system \cite{DGamma}; 
it may allow studies of CP violation with ``untagged'' $B_s$ data samples, 
where one does not distinguish between initially, i.e.\ at time $t=0$, 
present $B_s^0$ or $\overline{B^0_s}$ mesons \cite{dunietz}. The time 
evolution of the observables corresponding to (\ref{obs1}) takes the 
following form:
\begin{equation}\label{At}
|A_f(t)|^2=\frac{1}{2}\left[R_{\rm L}^f\,e^{-\Gamma_{\rm L}^{(q)}t}
+R_{\rm H}^f\,e^{-\Gamma_{\rm H}^{(q)}t}
+2\,e^{-\Gamma_q t}\left\{A_{\rm D}^f\cos(\Delta M_qt) + A_{\rm M}^f
\sin(\Delta M_qt)\right\}\right]
\end{equation}
\begin{equation}\label{Atbar}
|\overline{A}_f(t)|^2=\frac{1}{2}\left[R_{\rm L}^f\,
e^{-\Gamma_{\rm L}^{(q)}t}+R_{\rm H}^f\,e^{-\Gamma_{\rm H}^{(q)}t}
-2\,e^{-\Gamma_q t}\left\{A_{\rm D}^f\cos(\Delta M_qt) + A_{\rm M}^f
\sin(\Delta M_qt)\right\}\right],
\end{equation}
where $\Delta M_q\equiv M_{\rm H}^{(q)}-M_{\rm L}^{(q)}>0$ denotes the 
mass difference between the $B_q$ mass eigenstates, and $\Gamma_q\equiv
\left[\Gamma_{\rm L}^{(q)}+\Gamma_{\rm H}^{(q)}\right]/2$. The quantities 
$R_{\rm L}^f$, $R_{\rm H}^f$, $A_{\rm D}^f$ and $A_{\rm M}^f$, which are 
not independent from one another and satisfy the relation
\begin{equation}\label{Obs-rel}
\left(A_{\rm D}^f\right)^2+\left(A_{\rm M}^f\right)^2=R_{\rm L}^fR_{\rm H}^f\,,
\end{equation}
are given by 
\begin{eqnarray}
R_{\rm L}^f&=&|{\cal N}_f|^2\Bigl[(1+\eta_f\cos\phi_q)\nonumber\\
&&-2\,b_f\cos\rho_f\left\{\cos\omega+\eta_f\cos(\phi_q+\omega)\right\}
+b_f^2\left\{1+\eta_f\cos(\phi_q+2\,\omega)\right\}\Bigr]\\
R_{\rm H}^f&=&|{\cal N}_f|^2\Bigl[(1-\eta_f\cos\phi_q)\nonumber\\
&&-2\,b_f\cos\rho_f\left\{\cos\omega-\eta_f\cos(\phi_q+\omega)\right\}+
b_f^2\left\{1-\eta_f\cos(\phi_q+2\,\omega)\right\}\Bigr]\\
A_{\rm D}^f&\equiv&=2\,|{\cal N}_f|^2\,b_f\sin\rho_f\,\sin\omega
\label{AD-expr}\\
A_{\rm M}^f&=&\eta_f\,|{\cal N}_f|^2
\left[\sin\phi_q-2\,b_f\cos\rho_f\,\sin(\phi_q+\omega)+
b_f^2\sin(\phi_q+2\,\omega)\right].\label{AM-expr}
\end{eqnarray}
Here the phase $\phi_q$ denotes the CP-violating weak 
$B^0_q$--$\overline{B^0_q}$ mixing phase:
\begin{equation}
\phi_q=\left\{
\begin{array}{cc}
2\beta&\mbox{\,for $q=d$}\\
-2\delta\gamma&\mbox{\,for $q=s$,}
\end{array}
\right.
\end{equation}
where $2\delta\gamma\approx0.03$ is tiny in the Standard Model 
because of a Cabibbo suppression of ${\cal O}(\lambda^2)$. This phase 
cancels in 
\begin{equation}\label{RN-def}
S_f\equiv\frac{1}{2}\left(|A_f(0)|^2+|\overline{A}_f(0)|^2\right)=
\frac{1}{2}\left(R_{\rm L}^f+R_{\rm H}^f\right)=|{\cal N}_f|^2
\left(1-2\,b_f\cos\rho_f\cos\omega+b^2_f\right).
\end{equation}
It is also interesting to note that there are no $\Delta M_q t$ terms present
in the ``untagged'' combination
\begin{equation}\label{sum}
|A_f(t)|^2+|\overline{A}_f(t)|^2=R_{\rm L}^f\,e^{-\Gamma_{\rm L}^{(q)}t}+
R_{\rm H}^f\,e^{-\Gamma_{\rm H}^{(q)}t},
\end{equation}
whereas
\begin{equation}\label{diff}
|A_f(t)|^2-|\overline{A}_f(t)|^2=2\,e^{-\Gamma_q t}\left[
A_{\rm D}^f\cos(\Delta M_qt) + A_{\rm M}^f \sin(\Delta M_qt)\right].
\end{equation}
Because of (\ref{Obs-rel}), each of the $|A_f(t)|^2$ or 
$|\overline{A}_f(t)|^2$ ($f\in\{0,\parallel,\perp\}$) terms of the 
$B_q\to X_1\,X_2$ angular distribution provides three independent 
observables, which we may choose as $A_{\rm D}^f$, $A_{\rm M}^f$ and $S_f$.

The time evolution of the interference terms (\ref{obs2}) is analogous
to (\ref{sum}) and (\ref{diff}). Let us first give the expressions for
the observables corresponding to (\ref{RN-def}):
\begin{eqnarray}
\lefteqn{R\equiv\frac{1}{2}\left[\Re\{A_0^\ast(0)A_\parallel(0)\}+
\Re\{\overline{A}_0^\ast(0)\overline{A}_\parallel(0)\}\right]=
|{\cal N}_0||{\cal N}_\parallel|\Bigl[\,\cos\Delta_{0,\parallel}}\nonumber\\
&&\mbox{}\quad\,-\left\{b_0\cos(\rho_0-\Delta_{0,\parallel})+b_\parallel
\cos(\rho_\parallel+\Delta_{0,\parallel})\right\}\cos\omega+
b_0\,b_\parallel\cos(\rho_0-\rho_\parallel-\Delta_{0,\parallel})
\Bigr]\label{Re-sum}\\
\lefteqn{I^f_{\rm D}\equiv\frac{1}{2}\left[\Im\{A_f^\ast(0)A_\perp(0)\}+
\Im\{\overline{A}_f^\ast(0)\overline{A}_\perp(0)\}\right]}\nonumber\\
&&\mbox{}\quad\,=|{\cal N}_f||{\cal N}_\perp|\Bigl[\,b_f\cos(\rho_f-
\Delta_{f,\perp})-b_\perp\cos(\rho_\perp+\Delta_{f,\perp})\Bigr]
\sin\omega,\label{Im-sum}
\end{eqnarray}
where the 
\begin{equation}\label{Delta-def}
\Delta_{\tilde f,f}\equiv\delta_f-\delta_{\tilde f}
\end{equation}
denote the differences of the CP-conserving strong phases of the amplitudes
${\cal N}_f\equiv e^{i\delta_f}|{\cal N}_f|$ and 
${\cal N}_{\tilde f}\equiv e^{i\delta_{\tilde f}}|{\cal N}_{\tilde f}|$.
On the other hand, the rate differences corresponding to (\ref{diff}) 
take the following form:
\begin{eqnarray}
\Re\{A_0^\ast(t)A_\parallel(t)\}-
\Re\{\overline{A}_0^\ast(t)\overline{A}_\parallel(t)\}&=&2\, 
e^{-\Gamma_q t}\Bigl[R_{\rm D}\cos(\Delta M_qt) + R_{\rm M} \sin(\Delta M_qt)
\Bigr]\qquad\mbox{}\label{Re-diff}\\
\Im\{A_f^\ast(t)A_\perp(t)\}-\Im\{\overline{A}_f^\ast(t)
\overline{A}_\perp(t)\}&=&2\,e^{-\Gamma_q t}\left[I_f
\cos(\Delta M_qt)-I_{\rm M}^f \sin(\Delta M_qt)\right],
\quad\mbox{}\label{Im-diff}
\end{eqnarray}
where
\begin{eqnarray}
R_{\rm D}&=&|{\cal N}_0||{\cal N}_\parallel|
\left[b_0\sin(\rho_0-\Delta_{0,\parallel})+b_\parallel
\sin(\rho_\parallel+\Delta_{0,\parallel})\right]\sin\omega\label{RD-expr}\\
R_{\rm M}&=&|{\cal N}_0||{\cal N}_\parallel|\Bigl[
\cos\Delta_{0,\parallel}\sin\phi_q-\left\{b_0
\cos(\rho_0-\Delta_{0,\parallel})+b_\parallel\cos(\rho_\parallel+
\Delta_{0,\parallel})\right\}\sin(\phi_q+\omega)\nonumber\\
&&+\,\,b_0\,b_\parallel\cos(\rho_0-\rho_\parallel-\Delta_{0,\parallel})
\sin(\phi_q+2\,\omega)\Bigr]\label{RM-expr}
\end{eqnarray}
and
\begin{eqnarray}
I_f&=&|{\cal N}_f||{\cal N}_\perp|\Bigl[
\sin\Delta_{f,\perp}+\left\{b_f\sin(\rho_f-\Delta_{f,\perp})-
b_\perp\sin(\rho_\perp+\Delta_{f,\perp})\right\}\cos\omega\nonumber\\
&&-\,\,b_f\,b_\perp\sin(\rho_f-\rho_\perp-\Delta_{f,\perp})
\Bigr]\label{I-expr}\\
I_{\rm M}^f&=&|{\cal N}_f||{\cal N}_\perp|\Bigl[
\cos\Delta_{f,\perp}\cos\phi_q-\left\{b_f
\cos(\rho_f-\Delta_{f,\perp})+b_\perp\cos(\rho_\perp+\Delta_{f,\perp})
\right\}\cos(\phi_q+\omega)\nonumber\\
&&+\,\,b_f\,b_\perp\cos(\rho_f-\rho_\perp-\Delta_{f,\perp})
\cos(\phi_q+2\,\omega)\Bigr].
\end{eqnarray}
Note that $f\in\{0,\parallel\}$ in (\ref{Im-sum}) and (\ref{Im-diff}).
The minus sign in the latter expression is due to the different CP
eigenvalues of $f\in\{0,\parallel\}$ and $f=\,\perp$. If we set 
``$\,0=\,\,\parallel\,$'' in (\ref{RD-expr}) and (\ref{RM-expr}), we get
expressions taking the same form as (\ref{AD-expr}) and (\ref{AM-expr}), 
which provides a nice cross check. The expressions given above generalize
those derived in \cite{ddf1} in two respects: they take into account
penguin contributions, and they allow for a sizeable value of the
$B^0_q$--$\overline{B^0_q}$ mixing phase $\phi_q$. In the discussion of
$B_s\to J/\psi\,\phi$ in \cite{ddf1}, it was assumed that $\phi_s$
is a small phase, and terms of ${\cal O}(\phi_s^2)$ were neglected. 

\begin{figure}
\centerline{
\epsfxsize=4.3truecm
\epsffile{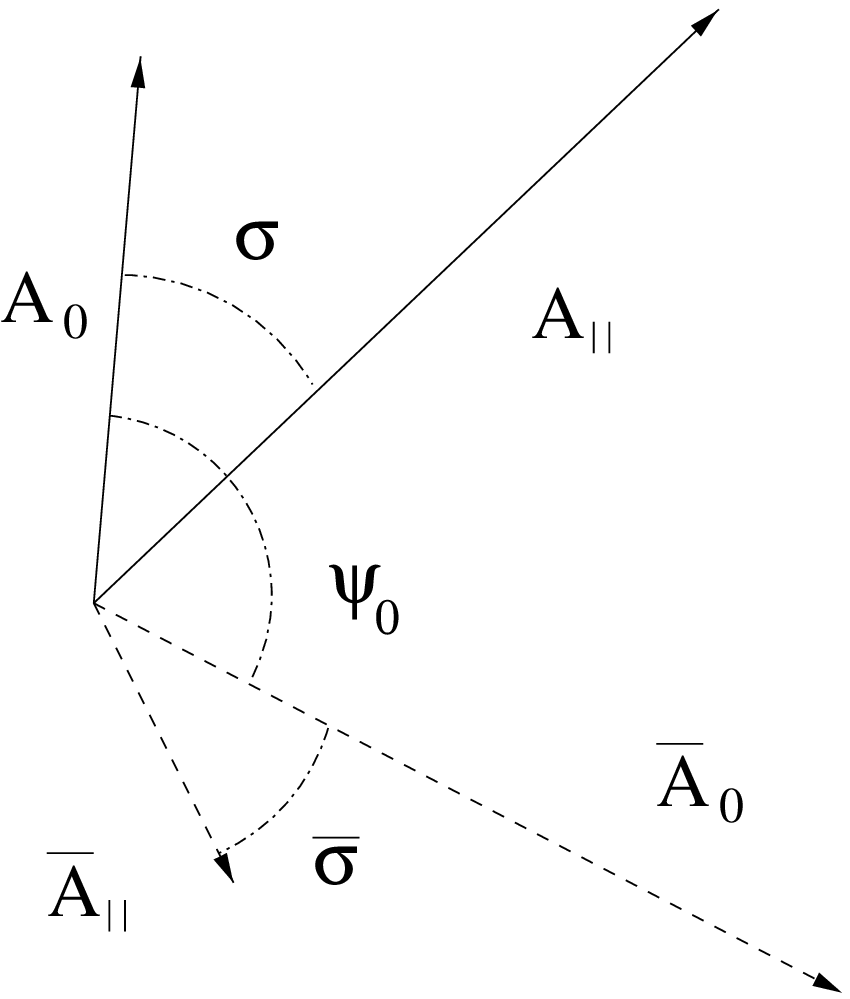}}
\caption{The amplitudes $A_0$, $A_\parallel$ and $\overline{A}_0$, 
$\overline{A}_\parallel$ in the complex plane.}\label{fig:ampls}
\end{figure}

Unfortunately, not all of the observables $S_f$, $A_{\rm D}^f$ and 
$A_{\rm M}^f$ are independent from those of the interference 
terms (\ref{obs2}). This can be seen by considering two different 
final-state configurations $f$ and $\tilde f$. In this case, the 
time-dependent angular distribution provides nine observables. To be 
definite, let us consider the case $f=0$ and $\tilde f=\,\parallel$. 
Then we have six observables, corresponding to $S_f$, 
$A_{\rm D}^f$ and $A_{\rm M}^f$ ($f\in\{0,\parallel\}$), as well as the 
three observables $R$, $R_{\rm D}$ and $R_{\rm M}$, which are due to the 
real parts in (\ref{obs2}). The measurement of $S_f$ and $A_{\rm D}^f$ 
allows us to fix the magnitudes $|A_0|$, $|A_\parallel|$ and 
$|\overline{A}_0|$, $|\overline{A}_\parallel|$. Using in addition the 
observables $R$ and $R_{\rm D}$, we can determine the angle $\sigma$ 
between the unmixed amplitudes $A_0$ and $A_\parallel$, as well as the 
angle $\overline{\sigma}$ between $\overline{A}_0$, $\overline{A}_\parallel$ 
(see Fig.\ \ref{fig:ampls}). So far, the relative orientation of the 
amplitudes $(A_0$, $A_\parallel)$ and $(\overline{A}_0$, 
$\overline{A}_\parallel)$ is not determined. However, if we use, in addition, 
the mixing-induced CP asymmetry $A_{\rm M}^0$, we are in a position to fix 
$\phi_q+\psi_0$, where $\psi_0$ denotes the angle between the amplitudes 
$A_0$ and $\overline{A}_0$:
\begin{equation}
A_{\rm M}^0=|A_0||\overline{A}_0|\sin(\phi_q+\psi_0).
\end{equation}
Since the relative orientation of the amplitudes $e^{-i\phi_q}
\overline{A}_\parallel$ and $A_\parallel$ is also fixed this way, we can 
predict the values of the two remaining mixing-induced CP-violating 
observables $A_{\rm M}^\parallel$ and $R_{\rm M}$. Consequently, only 
seven of the nine observables are independent from one another. 

It is convenient to introduce the following ``normalized'' observables:
\begin{equation}\label{A-hat}
\hat A_{\rm D}^f\equiv\frac{A_{\rm D}^f}{S_f},\quad
\hat A_{\rm M}^f\equiv\frac{A_{\rm M}^f}{S_f},
\end{equation}
\begin{equation}\label{R-hat}
\hat R\equiv\frac{R}{\sqrt{S_0S_\parallel}},\quad 
\hat R_{\rm D}\equiv\frac{R_{\rm D}}{\sqrt{S_0S_\parallel}},\quad 
\hat R_{\rm M}\equiv\frac{R_{\rm M}}{\sqrt{S_0S_\parallel}},
\end{equation}
\begin{equation}\label{I-hat}
\hat I_f\equiv\frac{I_f}{\sqrt{S_fS_\perp}},\quad 
\hat I^f_{\rm D}\equiv\frac{I^f_{\rm D}}{\sqrt{S_fS_\perp}},\quad
\hat I^f_{\rm M}\equiv\frac{I^f_{\rm M}}{\sqrt{S_fS_\perp}},
\end{equation}
which have the advantage that they do not depend on the overall normalization 
factors $|{\cal N}_f|$. The observables $\hat A_{\rm D}^f$ and 
$\hat A_{\rm M}^f$ allow us to determine the hadronic parameters $b_f$ and 
$\rho_f$ as functions of $\omega$ and $\phi_q$:
\begin{equation}\label{Bdet}
b_f=\sqrt{\frac{1}{k_f}\left[l_f\pm\sqrt{l_f^2-h_f k_f}\,\right]}
\end{equation}
\begin{equation}\label{Bcos}
2\,b_f\cos\rho_f=u_f+v_f\,b_f^2
\end{equation}
\begin{equation}\label{Bsin}
2\,b_f\sin\rho_f=\left[(1-u_f\cos\omega)+(1-v_f\cos\omega)\,b^2_f
\right]\left(\frac{\hat A_{\rm D}^f}{\sin\omega}\right),
\end{equation}
where
\begin{eqnarray}
h_f&=&u_f^2+D_f\,(1-u_f\cos\omega)^2\\
k_f&=&v_f^2+D_f\,(1-v_f\cos\omega)^2\\
l_f&=&2-u_f\,v_f-D_f\,(1-u_f\cos\omega)(1-v_f\cos\omega),
\end{eqnarray}
with 
\begin{eqnarray}
u_f&=&\frac{\eta_f\hat A_{\rm M}^f-\sin\phi_q}{
\eta_f\hat A_{\rm M}^f\cos\omega-\sin(\phi_q+\omega)}\label{u-def}\\
v_f&=&\frac{\eta_f\hat A_{\rm M}^f-\sin(\phi_q+2\,\omega)}{\eta_f
\hat A_{\rm M}^f\cos\omega-\sin(\phi_q+\omega)}\label{v-def}
\end{eqnarray}
and
\begin{equation}
D_f=\left(\frac{\hat A_{\rm D}^f}{\sin\omega}\right)^2.
\end{equation}
It should be emphasized that no approximations were made in order to
derive these expressions. If we consider, in addition to $\hat A_{\rm D}^f$ 
and $\hat A_{\rm M}^f$, either the observables specified in (\ref{R-hat}) 
or those given in (\ref{I-hat}), we obtain seven normalized observables, 
which depend on five hadronic parameters ($b_f$, $\rho_f$, $b_{\tilde f}$, 
$\rho_{\tilde f}$ and $\Delta_{\tilde f,f}$), as well as on the two 
CP-violating weak phases $\phi_q$ and $\omega$. However, only five of the 
seven observables are independent from one another, so that we have not 
sufficient observables at our disposal to extract these parameters 
simultaneously. To accomplish this task, we have to make use, for example, 
of another decay that can be related to $B_q\to X_1\,X_2$ through 
flavour-symmetry arguments. On the other hand, if we use $\phi_q$ and 
$\omega$ as an input, all hadronic parameters describing the decay 
$B_q\to X_1\,X_2$ can be extracted {\it without} any additional assumption, 
thereby providing valuable insights into hadronic physics and a very fertile 
testing ground for model calculations of $B_q\to X_1\,X_2$. The measurement 
of the angular distributions discussed in this paper requires high 
statistics and can probably only be performed at ``second-generation'' 
$B$-physics experiments at hadron machines, such as LHCb or BTeV, where 
also decays of $B_s$-mesons can be studied. Since several promising 
strategies to extract the weak phases $\phi_q$ and $\omega$ at such 
experiments were already proposed (see, for example, \cite{revs}), it 
may indeed be an interesting alternative to use measurements of angular 
distributions not to extract CKM phases, but to explore hadronic physics. 

In practical applications, the parameters $b_f$ typically measure the 
ratio of ``penguin'' to ``tree'' contributions. Applying the 
Bander--Silverman--Soni mechanism \cite{bss}, and following the formalism 
developed in \cite{pen-calc,kraetal}, which makes use -- among other 
things -- of the ``factorization hypothesis'', we obtain for various classes 
of $B$ decays
\begin{equation}\label{bss-rel}
b_f\equiv b,\quad\rho_f\equiv\rho\quad \forall\, f\in\{0,\parallel,\perp\}.
\end{equation}
The main reason for these relations is that the form factors, which
depend on the final-state configuration $f$, cancel in the ratios $b_f$
of ``penguin'' to ``tree'' contributions. Although non-factorizable 
contributions are expected to play an important role, thereby 
affecting (\ref{bss-rel}), it is interesting to investigate the 
implications of these relations on the observables of the angular 
distributions in more detail. If we introduce
\begin{eqnarray}
\hat A_{\rm D}&\equiv&=\frac{2\,b\,\sin\rho\,\sin\omega}{1-2\,b\,\cos\rho\,
\cos\omega+b^2}\\
\hat A_{\rm M}&=&\frac{\sin\phi_q-2\,b\,\cos\rho\,\sin(\phi_q+\omega)+
b^2\sin(\phi_q+2\,\omega)}{1-2\,b\,\cos\rho\,\cos\omega+b^2},
\end{eqnarray}
we obtain
\begin{equation}\label{A-bss}
\hat A_{\rm D}^f=\hat A_{\rm D},\quad
\hat A_{\rm M}^f=\eta_f\hat A_{\rm M},
\end{equation}
\begin{equation}\label{R-bss}
\hat R=\cos\Delta_{0,\parallel},\quad \hat R_{\rm D}=\hat A_{\rm D}
\cos\Delta_{0,\parallel},\quad \hat R_{\rm M}=\hat A_{\rm M}
\cos\Delta_{0,\parallel},
\end{equation}
\begin{equation}\label{I-bss}
\hat I_f=\sin\Delta_{f,\perp}\,\quad
\hat I_{\rm D}^f=\hat A_{\rm D}\sin\Delta_{f,\perp},
\end{equation}
\begin{equation}\label{IM-bss}
I\equiv\frac{\hat I_{\rm M}^f}{\cos\Delta_{f,\perp}}=
\left[\frac{\cos\phi_q-2\,b\,\cos\rho\,\cos(\phi_q+\omega)
+b^2\cos(\phi_q+2\,\omega)}{1-2\,b\,\cos\rho\,\cos\omega+b^2}\right],
\end{equation}
where
\begin{equation}
\left(\hat A_{\rm D}\right)^2+\left(\hat A_{\rm M}\right)^2+I\,^2=1.
\end{equation}
These relations provide an interesting test of whether (\ref{bss-rel}) 
is realized in the decay $B_q\to X_1\,X_2$. Note that $\hat I_{\rm M}^f$ 
does not -- in contrast to (\ref{I-bss}) -- vanish for trivial values of 
$\Delta_{f,\perp}$.

\boldmath
\section{Extracting CKM Phases and Hadronic Parameters}\label{STRAT}
\unboldmath
Let us now focus on the extraction of CKM phases from the observables of 
the $B_q\to X_1\,X_2$ angular distribution. As we have already noted, 
to this end, we have to employ an additional input, since we have only 
five independent normalized observables at our disposal, which depend 
on seven ``unknowns''. Although it would be desirable to determine 
$\phi_q$ and $\omega$ simultaneously, usually only the CKM phase $\omega$ 
is of central interest. 

The $B^0_s$--$\overline{B^0_s}$ mixing phase 
$\phi_s\equiv-2\delta\gamma=2\,\mbox{arg}(V_{ts}^\ast V_{tb})$ is 
negligibly small in the Standard Model. It can be probed -- and in principle
even determined -- with the help of the decay $B_s\to J/\psi\,\phi$ (see, 
for example, \cite{ddf1}). Large CP-violating effects in this decay would 
signal that $2\delta\gamma$ is not tiny, and would be a strong indication 
for new-physics contributions to $B^0_s$--$\overline{B^0_s}$ mixing. On
the other hand, the $B^0_d$--$\overline{B^0_d}$ mixing phase $\phi_d=2\beta$ 
can be fixed in a reliable way through the ``gold-plated'' mode $B_d\to 
J/\psi\,K_{\rm S}$ \cite{bisa}. Strictly speaking, mixing-induced 
CP violation in $B_d\to J/\psi\, K_{\rm S}$ probes $\sin(2\beta+\phi_K)$, 
where $\phi_K$ is related to the weak $K^0$--$\overline{K^0}$ mixing phase 
and is negligibly small in the Standard Model. Because of the small value 
of the CP-violating parameter $\varepsilon_K$ of the neutral kaon system, 
$\phi_K$ can only be affected by very contrived models of new physics 
\cite{nirsil}. A measurement of mixing-induced CP violation in 
$B_d\to J/\psi\,K_{\rm S}$ allows us to fix $\phi_d=2\beta$ only up to a 
twofold ambiguity. Several strategies to resolve this ambiguity were 
proposed in the literature \cite{ambig}, which should be feasible for 
``second-generation'' $B$-physics experiments. As we will see in the 
following section, also the decay $B_d\to J/\psi\,\rho^0$, in combination 
with $B_s\to J/\psi\,\phi$, allows us to accomplish this task.

If we use $\phi_q$ thus determined as an input and consider, in addition to 
$\hat A_{\rm D}^f$ and $\hat A_{\rm M}^f$, either the observables specified 
in (\ref{R-hat}) or those given in (\ref{I-hat}), we can determine $\omega$ 
as a function of a {\it single} hadronic parameter. Let us, for the moment, 
focus on the latter case, i.e.\ on the observables $\hat A_{\rm D}^f$, 
$\hat A_{\rm M}^f$, $\hat A_{\rm D}^\perp$, $\hat A_{\rm M}^\perp$ and 
$\hat I_f$, $\hat I_{\rm D}^f$, $\hat I_{\rm M}^f$ for a given final-state 
configuration $f\in\{0,\parallel\}$. Since $|{\cal N}_f|$ and 
$|{\cal N}_\perp|$ cancel in these quantities, they depend only on the 
hadronic parameters $b_f$, $\rho_f$, $b_\perp$, $\rho_\perp$, 
$\Delta_{f,\perp}$, as well as on the weak phases $\omega$ and $\phi_q$. 
Consequently, we have seven observables at our disposal, which depend on 
seven ``unknowns''. However, only five of the seven observables are
independent from one another, as we have discussed in the previous section.
If we use $\phi_q$ as an input, we can, for instance, obtain $\omega$ and 
$b_f$, $\rho_f$, $b_\perp$, $\rho_\perp$ as functions of the strong phase 
difference $\Delta_{f,\perp}$ in a {\it theoretically clean} way. Although
the following discussion deals with $\Delta_{f,\perp}$, we can also replace 
this quantity by another hadronic parameter of our choice. If we fix 
$\Delta_{f,\perp}$, for example, by comparing $B_q\to X_1\,X_2$ with an 
$SU(3)$-related mode, all parameters can be extracted. Using in addition 
the observables $S_f$, we can also determine the normalization 
factors $|N_f|$. Comparing them with those of the $SU(3)$-related mode used 
to fix $\Delta_{f,\perp}$, we can obtain valuable insights into 
$SU(3)$-breaking corrections. The observables that we have not used so far 
can be used to resolve discrete ambiguities, arising typically in the 
extraction of these parameters. 

Let us now give the formulae to implement this approach in a mathematical
way. The general expression for the observable $\hat I_f$ (see 
(\ref{I-expr}) and (\ref{I-hat})) leads to the equation
\begin{equation}\label{Del-eq}
A_f\sin\Delta_{f,\perp}+B_f\cos\Delta_{f,\perp}=C_f,
\end{equation}
where
\begin{eqnarray}
A_f&=&\frac{1}{N_f}\Bigl[1-\left(b_f\cos\rho_f+b_\perp\cos\rho_\perp\right)
\cos\omega+b_fb_\perp\left(\cos\rho_f\cos\rho_\perp+
\sin\rho_f\sin\rho_\perp\right)\Bigr]\\
B_f&=&\frac{1}{N_f}\Bigl[\left(b_f\sin\rho_f-b_\perp\sin\rho_\perp
\right)\cos\omega-b_fb_\perp\left(\sin\rho_f\cos\rho_\perp-
\cos\rho_f\sin\rho_\perp\right)\Bigr]\\
C_f&=&\hat I_f,
\end{eqnarray}
with
\begin{equation}
N_f=\sqrt{\Bigl(1-2\,b_f\cos\rho_f\cos\omega+b^2_f\Bigr)
\Bigl(1-2\,b_\perp\cos\rho_\perp\cos\omega+b^2_\perp\Bigr)}\,.
\end{equation}
The solution of (\ref{Del-eq}) is straightforward, and is 
given as follows:
\begin{equation}\label{SOL}
\sin\Delta_{f,\perp}=\frac{A_fC_f\pm\sqrt{\left(A_f^2+B_f^2-C_f^2\right)
B_f^2}}{A_f^2+B_f^2},\quad\cos\Delta_{f,\perp}=\frac{C_f-A_f
\sin\Delta_{f,\perp}}{B_f}.
\end{equation}
If we insert $b_f$ and $\rho_f$, determined as functions of $\omega$ and 
$\phi_q$ with the help of (\ref{Bdet})--(\ref{Bsin}), into the expressions
given above, we can -- for a given value of $\phi_q$ -- determine 
$\Delta_{f,\perp}$ as a function of $\omega$. It should be emphasized that
the relation between $\Delta_{f,\perp}$, $\omega$ and $\phi_q$ obtained
this way is valid {\it exactly}. Using $\hat I_{\rm D}^f$ or 
$\hat I_{\rm M}^f$ instead of $\hat I_f$ would lead to the same relation, 
since these observables are not independent from $\hat I_f$. 

Alternatively, we may use the observables (\ref{R-hat}) instead of 
(\ref{I-hat}). The general expression for $\hat R$ (see (\ref{Re-sum}) and
(\ref{R-hat})) implies an equation similar to (\ref{Del-eq}), where $A_f$, 
$B_f$ and $C_f$ have to be replaced through
\begin{eqnarray}
A&=&\frac{1}{N}\Bigl[\left(b_\parallel\sin\rho_\parallel-
b_0\sin\rho_0\right)\cos\omega+b_0b_\parallel\left(\sin\rho_0
\cos\rho_\parallel-\cos\rho_0\sin\rho_\parallel\right)\Bigr]\\
B&=&\frac{1}{N}\Bigl[1-\left(b_0\cos\rho_0+b_\parallel\cos\rho_\parallel
\right)\cos\omega+b_0b_\parallel\left(\cos\rho_0\cos\rho_\parallel+
\sin\rho_0\sin\rho_\parallel\right)\Bigr]\\
C&=&\hat R,
\end{eqnarray}
where
\begin{equation}
N=\sqrt{\Bigl(1-2\,b_0\cos\rho_0\cos\omega+b^2_0\Bigr)
\Bigl(1-2\,b_\parallel\cos\rho_\parallel\cos\omega+b^2_\parallel\Bigr)}\,.
\end{equation}
Obviously, the most efficient strategy of combining the observables
provided by the $B_q\to X_1\,X_2$ angular distribution depends on their
actually measured values. 

If we are willing to make more extensive use of flavour-symmetry arguments 
than just to fix the strong phase difference $\Delta_{\tilde f,f}$, it is 
in principle possible to determine also the $B^0_q$--$\overline{B^0_q}$ 
mixing phase $\phi_q$. In the following section, we will have a closer 
look at the decay $B_d\to J/\psi\,\rho^0$, which can be related to  
$B_s\to J/\psi\,\phi$ through $SU(3)$ arguments and a certain dynamical 
assumption concerning ``exchange'' and ``penguin annihilation'' topologies. 
However, before we turn to these modes, which allow the simultaneous 
extraction of $\phi_d=2\beta$ and $\gamma$, let us first give two useful 
expressions for the observables $\hat R$ and $\hat I_f$. Since the parameters 
$b_f$ measure typically the importance of ``penguin'' topologies in comparison 
with current--current contributions, they may not be too large. If we 
eliminate the hadronic parameters $b_f$ and $\rho_f$ in $\hat R$ and 
$\hat I_f$ with the help of the observables $\hat A_{\rm D}^f$ and 
$\hat A_{\rm M}^f$ and keep only the leading-order terms in $b_f$, we 
obtain
\begin{eqnarray}
\hat R&\approx&\cos\Delta_{0,\parallel}-\frac{1}{2}\left(\hat A_{\rm D}^0-
\hat A_{\rm D}^\parallel\right)\frac{\sin\Delta_{0,\parallel}}{\tan\omega}
\label{R-approx}\\
\hat I_f&\approx&\sin\Delta_{f,\perp}+\frac{1}{2}\left(
\hat A_{\rm D}^f-\hat A_{\rm D}^\perp\right)
\frac{\cos\Delta_{f,\perp}}{\tan\omega},\label{ID-approx}
\end{eqnarray}
allowing us to determine $\omega$ if the strong phase differences 
$\Delta_{0,\parallel}$ or $\Delta_{f,\perp}$ are known. Interestingly,
the leading-order expressions (\ref{R-approx}) and (\ref{ID-approx})
do not depend on the $B^0_q$--$\overline{B^0_q}$ mixing phase $\phi_q$. 
A possible disadvantage of $\hat R$ is that $\omega$ enters in combination 
with $\sin\Delta_{0,\parallel}$. Since $\Delta_{0,\parallel}$ is a difference 
of CP-conserving strong phases (see (\ref{Delta-def})), it may be small, 
thereby weakening the sensitivity of these observables on $\omega$. The 
situation concerning this point is very different in the case of the 
observables $\hat I_f$, which allow us to determine $\omega$ even in the 
case of $\Delta_{f,\perp}\in\{0^\circ,180^\circ\}$.

\boldmath
\section{Extracting $\beta$ and $\gamma$ from  $B_d\to J/\psi\,\rho^0$ and\\ 
$B_s\to J/\psi\,\phi$}\label{BPSI}
\unboldmath
If we combine the observables describing the time-dependent angular 
distribution of the decay $B_d\to J/\psi[\to l^+l^-]\,
\rho^0[\to\pi^+\pi^-]$ with those of $B_s\to J/\psi[\to l^+l^-]\,
\phi[\to K^+K^-]$, we may extract the $B^0_d$--$\overline{B^0_d}$ mixing 
phase $\phi_d=2\beta$ and the angle $\gamma$ of the unitarity triangle. 
The $B_d\to J/\psi\,\rho^0$ angular distribution can be obtained 
straightforwardly from the $B_s\to J/\psi\,\phi$ case, which has been 
discussed in detail in \cite{ddf1}, by performing appropriate replacements 
of kinematical variables. 

The decay $B_d^0\to J/\psi\,\rho^0$ originates from $\bar b\to\bar cc\bar d$
quark-level transitions; the structure of its decay amplitude is 
completely analogous to the one of $B_s^0\to J/\psi\,K_{\rm S}$ (see
\cite{RF1}). For a given final-state configuration $f$ with CP eigenvalue 
$\eta_f$, we have
\begin{equation}\label{Bd-ampl1}
A(B_d^0\to [J/\psi\, \rho^0]_f)=\lambda_c^{(d)}\left[A_{\rm cc}^{(c)f}+
A_{\rm pen}^{(c)f}\right]+\lambda_u^{(d)}A_{\rm pen}^{(u)f}
+\lambda_t^{(d)}A_{\rm pen}^{(t)f}\,,
\end{equation}
where $A_{\rm cc}^{(c)f}$ is due to current--current contributions, and 
the amplitudes $A_{\rm pen}^{(q)f}$ describe penguin topologies with internal 
$q$ quarks ($q\in\{u,c,t\})$. These penguin amplitudes take into account 
both QCD and electroweak penguin contributions, and
\begin{equation}
\lambda_q^{(d)}\equiv V_{qd}V_{qb}^\ast
\end{equation}
are the usual CKM factors. Employing the unitarity of the CKM matrix and 
the Wolfenstein parametrization \cite{wolf}, generalized to 
include non-leading terms in $\lambda$ \cite{blo}, we obtain
\begin{equation}\label{Bd-ampl}
A(B_d^0\to [J/\psi\, \rho^0]_f)=-\lambda\,{\cal A}_f\left[1-a_f\,e^{i\theta_f}
e^{i\gamma}\right],
\end{equation}
where
\begin{equation}\label{Aa-def}
{\cal A}_f\equiv\lambda^2A\left[A_{\rm cc}^{(c)f}+A_{\rm pen}^{(ct)f}\right],
\end{equation}
with $A_{\rm pen}^{(ct)f}\equiv A_{\rm pen}^{(c)f}-A_{\rm pen}^{(t)f}$, and
\begin{equation}\label{a-def}
a_f\,e^{i\theta_f}\equiv R_b\left(1-\frac{\lambda^2}{2}\right)\left[
\frac{A_{\rm pen}^{(ut)f}}{A_{\rm cc}^{(c)f}+A_{\rm pen}^{(ct)f}}\right].
\end{equation}
The quantity $A_{\rm pen}^{(ut)f}$ is defined in analogy to 
$A_{\rm pen}^{(ct)f}$, and the relevant CKM factors are given as follows:
\begin{equation}\label{CKM-exp}
\lambda\equiv|V_{us}|=0.22\,,\quad A\equiv\frac{1}{\lambda^2}
\left|V_{cb}\right|=0.81\pm0.06\,,\quad R_b\equiv\frac{1}{\lambda}
\left|\frac{V_{ub}}{V_{cb}}\right|=0.41\pm0.07\,.
\end{equation}
It should be emphasized that (\ref{Bd-ampl}) is a completely general 
parametrization of the $B_d^0\to J/\psi\,\rho^0$ decay amplitude within the 
Standard Model, relying only on the unitarity of the CKM matrix. In 
particular, this expression takes also into account final-state-interaction 
effects, which can be considered as long-distance penguin topologies with 
internal up- and charm-quark exchanges \cite{bfm}. Comparing 
(\ref{Bd-ampl}) with (\ref{ampl}), we observe that 
\begin{equation}
{\cal N}_f=-\lambda\,{\cal A}_f,\quad b_f=a_f,\quad \rho_f=\theta_f,\quad
\omega=\gamma.
\end{equation}

Let us now turn to $B_s^0\to J/\psi\,\phi$. Using the same notation as 
in (\ref{Bd-ampl}), we have
\begin{equation}\label{Bd-ampl2}
A(B_s^0\to [J/\psi\,\phi]_f)=\left(1-\frac{\lambda^2}{2}\right){\cal A}_f'
\left[1+\epsilon\,a_f'\,e^{i\theta_f'}
e^{i\gamma}\right],
\end{equation}
where ${\cal A}_f'$ and $a_f'e^{i\theta_f'}$ take the same form as
(\ref{Aa-def}) and (\ref{a-def}), respectively, and 
\begin{equation}
\epsilon\equiv\frac{\lambda^2}{1-\lambda^2}\,.
\end{equation}
The primes remind us that we are dealing with a $\bar b\to\bar s$ 
transition. Consequently, if we compare (\ref{Bd-ampl2}) with
(\ref{ampl}), we obtain
\begin{equation}
{\cal N}_f=\left(1-\frac{\lambda^2}{2}\right){\cal A}_f',\quad
b_f=\epsilon\,a_f',\quad\rho_f=\theta_f'+180^\circ,\quad\omega=\gamma.
\end{equation}

The $B_s\to J/\psi\,\phi$ and $B_d\to J/\psi\,\rho^0$ observables
can be related to each other through
\begin{equation}\label{Rel-1}
|{\cal A}_f'|=\sqrt{2}\,|{\cal A}_f|
\end{equation}
\begin{equation}\label{Rel-1p}
\Delta_{\tilde f,f}'=\Delta_{\tilde f,f}
\end{equation}
\begin{equation}\label{Rel-2}
a_f'= a_f,\quad \theta_f'=\theta_f,
\end{equation}
where the factor of $\sqrt{2}$ is due to the $\rho^0$ wave function. 
These relations rely both on the $SU(3)$ flavour symmetry of strong
interactions and on the neglect of certain ``exchange'' and ``penguin 
annihilation'' topologies. Although such topologies, which can be probed,
for example, through $B_s\to\rho^+\rho^-$, $D^{\ast+}D^{\ast-}$ decays, are 
usually expected to play a very minor role, they may in principle be enhanced 
through final-state-interaction effects \cite{FSI}. For the following 
considerations, it is useful to introduce the quantities
\begin{equation}\label{H-def}
H_f\equiv\frac{1}{\epsilon}\left(\frac{|{\cal A}_f'|}{|{\cal A}_f|}\right)^2
\frac{S_f}{S_f'}=\frac{1-2\,a_f\,\cos\theta_f\cos\gamma+a_f^2}{1+
2\,\epsilon\,a_f'\cos\theta_f'\cos\gamma+\epsilon^2\,a_f'^2}\,,
\end{equation}
which can be fixed through the ``untagged'' $B_d\to J/\psi\,\rho^0$ and 
$B_s\to J/\psi\,\phi$ observables with the help of (\ref{Rel-1}). 
Consequently, each of the linear polarization states 
$f\in\{0,\parallel,\perp\}$ provides the following three observables:
\begin{equation}\label{U-obs}
H_f,\,\,\hat A_{\rm D}^f,\,\, \hat A_{\rm M}^f.
\end{equation}
Applying (\ref{Rel-2}) to (\ref{H-def}), these observables depend only on 
the hadronic parameters $a_f$ and $\theta_f$, as well as on the 
$B^0_d$--$\overline{B^0_d}$ mixing phase $\phi_d=2\beta$ and the angle
$\gamma$ of the unitarity triangle. If we choose two different linear 
polarization states, the observables (\ref{U-obs}) allow us to determine 
the corresponding hadronic parameters and $\beta$ and $\gamma$ 
simultaneously. 

This approach can be implemented in a mathematical way as follows: 
if we consider a given final-state configuration $f$ and 
combine the observables $H_f$ and $\hat A_{\rm D}^f$, which do not 
depend on $\phi_d$, with each other, we can determine $a_f$ and $\theta_f$ 
as functions of~$\gamma$:
\begin{equation}\label{a-expr}
a_f=\sqrt{p_f\pm\sqrt{p_f^2-q_f}}
\end{equation}
\begin{equation}
2\,a_f\,\cos\theta_f=\frac{1-H_f+(1-\epsilon^2H_f)\,a_f^2}{(1+\epsilon H_f)
\cos\gamma}
\end{equation}
\begin{equation}
2\,a_f\,\sin\theta_f=\left[\frac{(1+\epsilon)(1+\epsilon\,a_f^2)H_f}{(1+
\epsilon H_f)}\right]\left(\frac{\hat A_{\rm D}^f}{\sin\gamma}\right),
\end{equation}
where
\begin{equation}
p_f=\frac{\left[2\left(1+\epsilon H_f\right)^2\cos^2\gamma-(1-H_f)\left(
1-\epsilon^2H_f\right)\right]\sin^2\gamma-\epsilon_f E_f}{\left(1-
\epsilon^2H_f\right)^2\sin^2\gamma+\epsilon^2 E_f}
\end{equation}
\begin{equation}
q_f=\frac{\left(1-H_f\right)^2\sin^2\gamma+E_f}{\left(1-\epsilon^2H_f\right)^2
\sin^2\gamma+\epsilon^2 E_f},
\end{equation}
with
\begin{equation}
E_f=\left[(1+\epsilon)\,H_f\,\hat A_{\rm D}^f\,\cos\gamma\right]^2.
\end{equation}
These expressions allow us to eliminate the hadronic parameters $a_f$ and
$\theta_f$ in the mixing-induced CP asymmetry $\hat A_{\rm M}^f$, thereby
fixing a contour in the $\gamma$--$\phi_d$ plane, which is related~to
\begin{equation}\label{phi-eq}
\tilde A_f\sin\phi_d+\tilde B_f\cos\phi_d=\tilde C_f,
\end{equation}
with
\begin{eqnarray}
\tilde A_f&=&1-2\,a_f\,\cos\theta_f\cos\gamma+a_f^2\cos2\gamma\\
\tilde B_f&=&-\,2\,a_f\,\cos\theta_f\sin\gamma+a_f^2\sin2\gamma\\
\tilde C_f&=&\left(1-2\,a_f\,\cos\theta_f\cos\gamma+a_f^2\right)(\eta_f
\hat A_{\rm M}^f).
\end{eqnarray}
The solution of (\ref{phi-eq}) has already been given in (\ref{SOL}). If we 
consider two different final-state configuratons $f$ and $\tilde f$, we 
obtain two different contours in the $\gamma$--$\phi_d$ plane; their 
intersection allows us to determine both $\gamma$ and $\phi_d=2\beta$. 
Using, in addition, the observables (\ref{R-hat}) or (\ref{I-hat}) -- 
depending on which final-state configurations $f$ and $\tilde f$ we 
consider -- we may resolve discrete ambiguities, arising typically in 
the extraction of $\phi_d$ and $\gamma$.

Because of the strong suppression of $a_f'$ through $\epsilon=0.05$ 
in (\ref{H-def}), this approach is essentially unaffected by possible 
corrections to (\ref{Rel-2}), and relies predominantly on the 
relation (\ref{Rel-1}). If we insert the values of $\phi_d$ and 
$\gamma$ thus determined into the expressions for the observables of the 
third linear polarization state $f'$, which has not been used so far, 
its hadronic parameters $|{\cal A}_{f'}|$, $a_{f'}$ and 
$\theta_{f'}$ can also be determined. Comparing $|{\cal A}_{f'}|$ 
with the $B_s\to J/\psi\,\phi$ parameter $|{\cal A}_{f'}'|$, we can
obtain valuable insights into the validity of (\ref{Rel-1}). Moreover, 
several other interesting cross checks can be performed with the many 
observables of the angular distributions. Because of our poor understanding 
of the hadronization dynamics of non-leptonic $B$ decays, only the 
``factorization'' approximation can be used for the time being to estimate 
factorizable $SU(3)$-breaking corrections to (\ref{Rel-1}). Explicit 
expressions for the $B_s\to J/\psi\,\phi$ observables can be found in 
\cite{ddf1}, and $SU(3)$-breaking effects in the corresponding form 
factors were studied in \cite{BaBr}. However, also non-factorizable effects 
are expected to play an important role, and experimental insights into 
these issues would be very helpful to find a better theoretical description. 

The simultaneous extraction of $\phi_d$ and $\gamma$ discussed above
works only if the hadronic parameters $a_f$ and $\theta_f$ are sufficiently
different from each other for two different final-state configurations $f$. 
If, for example, (\ref{bss-rel}) should apply to $B_d\to J/\psi\,\rho^0$ -- 
which seems to be quite unlikely -- the $B^0_d$--$\overline{B^0_d}$ mixing 
phase has to be fixed separately in order to determine $\gamma$. In this 
case, each linear polarization state $f\in\{0,\parallel,\perp\}$ provides 
a strategy to extract $\gamma$ that is completely analogous to the one 
proposed in \cite{RF1}, which makes use of $B_{s(d)}\to J/\psi\,K_{\rm S}$ 
decays. If we combine $H_f$ with $\hat A_{\rm M}^f$, we obtain
\begin{equation}
a_f=\sqrt{\frac{H_f-1+u_f\,(1+\epsilon H_f)\cos\gamma}{1-v_f\,(1+
\epsilon H_f)\cos\gamma-\epsilon^2 H_f}}\,.
\end{equation}
The intersection of the contours in the $\gamma$--$a_f$ plane described by
this expression with those related to (\ref{Bdet}) allows us to determine 
$\gamma$ and $a_f$. 

If we use $\phi_d$ as an input in order to extract $\gamma$ from 
$B_d\to J/\psi\,\rho^0$, it is, however, more favourable to follow the 
approach discussed in the previous section, i.e.\ to use (\ref{SOL}), and 
to fix $\Delta_{f,\perp}$ (or $\Delta_{0,\parallel}$) through the 
the $B_s\to J/\psi\,\phi$ observables with the help of (\ref{Rel-1p})
and (\ref{Rel-2}). Using in addition the observables involving the third 
linear polarization state $f'$ that we have not employed so far, we can 
also fix its hadronic parameters $a_{f'}$ and $\theta_{f'}$, as well as
the strong phase difference $\Delta_{f',f}$. Comparing $\Delta_{f',f}$ 
with its $B_s\to J/\psi\,\phi$ counterpart $\Delta_{f',f}'$, we may obtain 
valuable insights into possible corrections to (\ref{Rel-1p}).

As an interesting by-product, this strategy allows us to take into account 
also the penguin effects in the extraction of the 
$B^0_s$--$\overline{B^0_s}$ mixing phase $\phi_s$ from $B_s\to J/\psi\,\phi$. 
Although the penguin contributions are strongly suppressed in this mode 
because of the tiny parameter $\epsilon=0.05$ (see (\ref{Bd-ampl2})),
they may well lead to uncertainties of the extracted value of $\phi_s$
at the level of $10\%$, since $\phi_s={\cal O}(0.03)$ within the 
Standard Model. A measurement of $\phi_s=-2\lambda^2\eta$ would allow
us to determine the Wolfenstein parameter $\eta$ \cite{wolf}, thereby fixing 
the height of the unitarity triangle. Since the decay $B_s\to J/\psi\,\phi$
is very accessible at ``second-generation'' $B$-physics experiments performed 
at hadron machines, for instance at LHCb, it is an important issue to think 
about the hadronic uncertainties affecting the determination of $\phi_s$ from 
the corresponding angular distribution. The approach discussed above allows 
us to control these uncertainties with the help of $B_d\to J/\psi\, \rho^0$.

The experimental feasibility of the determination of $\gamma$ from the
$B_d\to J/\psi\,\rho^0$ angular distribution depends strongly on the 
``penguin parameters'' $a_f$. It is very difficult to estimate these 
quantities theoretically. In contrast to the ``usual'' QCD penguin 
topologies, the QCD penguins contributing to $B_d\to J/\psi\, \rho^0$ 
require a colour-singlet exchange, i.e.\ are ``Zweig-suppressed''. Such 
a comment does not apply to the electroweak penguins, which contribute in 
``colour-allowed'' form. The current--current amplitude $A_{\rm cc}^{(c)f}$ 
originates from ``colour-suppressed'' topologies, and the ratio 
$A_{\rm pen}^{(ut)f}/\left[A_{\rm cc}^{(c)f}+A_{\rm pen}^{(ct)f}\right]$, 
which governs $a_f$, may be sizeable. It would be very important to have 
a better theoretical understanding of the quantities $a_f\,e^{i\theta_f}$. 
However, such analyses are far beyond the scope of this paper, and are 
left for further studies. 

If the parameters $a_f$ should all be very small, which would be indicated
by $A_{\rm D}^f=R_{\rm D}=I_{\rm D}^f=0$, we could still determine the 
$B^0_d$--$\overline{B^0_d}$ mixing phase from the observables 
$\hat A_{\rm M}^f=\eta_f\,\sin\phi_d$. If we use, in addition, 
$\hat I_{\rm M}^f=\cos\Delta_{f,\perp}\cos\phi_d$ and fix 
$\cos\Delta_{f,\perp}$ through the corresponding $B_s\to J/\psi\,\phi$
observable, $\cos\phi_d$ can be determined as well. Consequently,
the $B^0_d$--$\overline{B^0_d}$ mixing phase $\phi_d$ can be fixed
{\it unambiguously} this way, thereby resolving a twofold ambiguity, which 
arises in the extraction of $\phi_d$ from $B_d\to J/\psi\,K_{\rm S}$. This 
mode probes only $\sin\phi_d$. Since $\phi_d=2\beta$, we are left with a 
twofold ambiguity for $\beta\in[0^\circ,360^\circ]$. If we assume that 
$\beta\in[0^\circ,180^\circ]$, as implied by the measured value of 
$\varepsilon_K$, we can fix $\beta$ unambiguously. For alternative methods 
to deal with ambiguities of this kind, see \cite{ambig}.

Before we turn to $B_d\to\rho \rho$ and $B_s\to K^{\ast}\,
\overline{K^{\ast}}$ decays, let us note that the approach presented in 
this section can also be applied to the angular distributions of the decay 
products of $B_{s(d)}\to J/\psi[\to l^+l^-]\,K^\ast[\to\pi^0K_{\rm S}]$ 
and $B_{d(s)}\to D^{\ast+}_{d(s)}\,D^{\ast-}_{d(s)}$. For the 
$B_{s(d)}\to J/\psi\,K_{\rm S}$ and $B_{d(s)}\to D^{+}_{d(s)} D^{-}_{d(s)}$ 
variants of these strategies, see \cite{RF1}.

\boldmath
\section{Further Applications}\label{APPL}
\unboldmath
In this section, we discuss further applications of the general strategies 
presented in Section~\ref{STRAT}. All of the methods discussed below have 
counterparts using $B_{d,s}$ decays into two pseudoscalar mesons. If we 
replace the pseudoscalars by higher resonances, for example, by vector 
mesons, as in the following discussion, the angular distributions of their 
decay products provide interesting alternative ways to extract CKM phases 
and hadronic parameters, going beyond the $B_{d,s}\to PP$ strategies. 
Because of the many observables provided by the angular distributions, we 
can, moreover, perform many interesting cross checks, for example, of 
certain flavour-symmetry relations.

\boldmath
\subsection{The Decays $B_d\to\rho^+\rho^-$ and 
$B_s\to K^{\ast+}K^{\ast-}$}\label{BdRhoRho}
\unboldmath
The decay $B_d^0\to\rho^+\rho^-$ originates from $\bar b\to\bar uu\bar d$ 
quark-level processes. Using the same notation as in (\ref{Bd-ampl}), we
have
\begin{equation}\label{Bdrhorho-ampl}
A(B_d^0\to[\rho^+\rho^-]_f)=\left(1-\frac{\lambda^2}{2}\right)
{\cal C}_f\,e^{i\gamma}\left[1-d_f\,e^{i\Theta_f}e^{-i\gamma}\right],
\end{equation}
where
\begin{equation}\label{Cf-def}
{\cal C}_f\equiv\lambda^3A\,R_b\left[\tilde A_{\rm cc}^{(u)f}+
\tilde A_{\rm pen}^{(ut)f}\right]
\end{equation}
and 
\begin{equation}\label{df-def}
d_f\,e^{i\Theta_f}\equiv\frac{1}{(1-\lambda^2/2)R_b}
\left[\frac{\tilde A_{\rm pen}^{(ct)f}}{\tilde A_{\rm cc}^{(u)f}+
\tilde A_{\rm pen}^{(ut)f}}\right].
\end{equation}
In order to distinguish the $B_d^0\to\rho^+\rho^-$ amplitudes from the
$B_d^0\to J/\psi\,\rho^0$ case discussed in the previous section, we
have introduced the tildes. The phase structure of the $B_d^0\to\rho^+\rho^-$
decay amplitude given in (\ref{Bdrhorho-ampl}), which is an exact
parametrization within the Standard Model, is completely analogous to 
the one for the $B_d^0\to\pi^+\pi^-$ amplitude given in \cite{RF2}, 
where a more detailed discussion can be found.

The expressions for the observables describing the time evolution of the 
angular distribution of the decay products of $B_d^0\to
\rho^+[\to\pi^+\pi^0]\,\,\rho^-[\to\pi^-\pi^0]$ can be obtained
straightforwardly from the formulae given in Section~\ref{EVOL}, by 
performing the following substitutions:
\begin{equation}
{\cal N}_f=\left(1-\frac{\lambda^2}{2}\right){\cal C}_f,\quad 
b_f=d_f,\quad \rho_f=\Theta_f,\quad \omega=-\,\gamma.
\end{equation}
Because of the factor of $e^{i\gamma}$ in front of the square brackets 
on the right-hand side of (\ref{Bdrhorho-ampl}), we have to deal with a 
small complication. Since the observables are governed by
\begin{equation}
\xi_f=e^{-i\phi_d}\frac{\overline{A}_f}{A_f}=\eta_f\, e^{-i(\phi_d+2\gamma)}
\left[\frac{1-d_f\,e^{i\Theta_f}e^{+i\gamma}}{1-d_f\,e^{i\Theta_f}e^{-i\gamma}}
\right],
\end{equation}
we have to do the following replacement, in addition:
\begin{equation}
\phi_d\to\phi_d+2\gamma\,.
\end{equation}

Let us now turn to the decay $B_s^0\to K^{\ast+}K^{\ast-}$, which is due 
to $\bar b\to\bar uu\bar s$ quark-level transitions. For a given
final-state configuration $f$ of the $K^{\ast+}K^{\ast-}$ pair, its 
decay amplitude can be parametrized as follows:
\begin{equation}\label{Bs-ampl}
A(B_s^0\to[K^{\ast+}K^{\ast-}]_f)=\lambda\,{\cal C}_f'\,e^{i\gamma}
\left[1+\left(\frac{1-\lambda^2}{\lambda^2}\right)d_f'\,e^{i\Theta_f'}
e^{-i\gamma}\right],
\end{equation}
where ${\cal C}_f'$ and $d_f'\,e^{i\Theta_f'}$ take the same form as 
(\ref{Cf-def}) and (\ref{df-def}), respectively, and the primes have been
introduced to remind us that we are dealing with a $\bar b\to\bar s$
mode. The phase structure of (\ref{Bs-ampl}) is completely analogous to 
the $B^0_s\to K^+K^-$ decay amplitude \cite{RF2}. The observables of the 
time-dependent $B_s\to K^{\ast+}[\to \pi K]\,K^{\ast-}[\to\overline{\pi}
\overline{K}]$ angular distribution can be obtained straightforwardly from 
the formulae given in Section~\ref{EVOL} by simply using the replacements:
\begin{equation}\label{BsKK-sub}
{\cal N}_f=\lambda\,{\cal C}_f',\quad 
b_f=\left(\frac{1-\lambda^2}{\lambda^2}\right)d_f',\quad 
\rho_f=\Theta_f'+180^\circ,\quad\omega=-\,\gamma.
\end{equation}
Moreover, we have to perform the substitution
\begin{equation}
\phi_s\to\phi_s+2\gamma
\end{equation}
because of the factor of $e^{i\gamma}$ in front of the square brackets 
in (\ref{Bs-ampl}). 

Explicit expressions for the $B_d\to\rho^+\rho^-$ and 
$B_s\to K^{\ast+}K^{\ast-}$ angular distributions in terms of 
helicity amplitudes can be found in \cite{kraetal}. Since 
$B_d^0\to\rho^+\rho^-$ and $B_s^0\to K^{\ast+}K^{\ast-}$ are related to 
each other by interchanging all down and strange quarks, the $U$-spin 
flavour symmetry of strong interactions implies
\begin{equation}\label{su3-1}
|{\cal C}_f'|=|{\cal C}_f|,\quad 
d_f'=d_f,\quad \Theta_f'=\Theta_f,
\end{equation}
as well as
\begin{equation}\label{suRel-1p}
\Delta_{\tilde f,f}'=\Delta_{\tilde f,f}.
\end{equation}
In contrast to (\ref{Rel-1})--(\ref{Rel-2}), these relations do
not rely on any dynamical assumption -- just on the $U$-spin flavour
symmetry. They can be used to combine the $B_d\to\rho^+\rho^-$ and 
$B_s\to K^{\ast+}K^{\ast-}$ observables with each other, thereby 
allowing the extraction of the CKM angle $\gamma$ and of the
$B^0_{d,s}$--$\overline{B^0_{d,s}}$ mixing phases $\phi_d=2\beta$ and
$\phi_s=-2\delta\gamma$. In contrast to the $B_d\to\pi^+\pi^-$, 
$B_s\to K^+K^-$ variant of this approach proposed in \cite{RF2},
{\it both} mixing phases and the CKM angle $\gamma$ can in principle 
be determined simultaneously. However, for the extraction of $\gamma$,
it is more favourable to fix $\phi_d$ and $\phi_s$ separately. Then
we are in a position to determine two contours in the 
$\gamma$--$\Delta_{\tilde f,f}$ and $\gamma$--$\Delta_{\tilde f,f}'$ 
planes in a {\it theoretically clean} way with the help of (\ref{SOL}). 
Using now the $U$-spin relation (\ref{suRel-1p}), $\gamma$ and all hadronic 
parameters describing the decays $B_d\to\rho^+\rho^-$ and 
$B_s\to K^{\ast+}K^{\ast-}$ can be determined. As we have already noted, 
the hadronic parameters provide a very fertile testing ground for model 
calculations of the decays $B_d\to\rho^+\rho^-$ and 
$B_s\to K^{\ast+}K^{\ast-}$. In particular, the penguin parameters 
$d_f\,e^{i\Theta_f}$ and $d_f'\,e^{i\Theta_f'}$ would be very interesting;
comparing their values with each other, we could obtain valuable insights 
into $U$-spin-breaking corrections. Moreover, there is one strong phase 
difference $\Delta_{f',f}$ left, which can be compared with its $U$-spin 
counterpart $\Delta'_{f',f}$. If we should find a small difference between 
these phases, it would be quite convincing to assume that our $U$-spin 
input (\ref{suRel-1p}) is also not affected by large corrections.

Let us finally note that there is another interesting way to parametrize 
the $B_d\to\rho^+\rho^-$ decay amplitudes (see also \cite{charles}). If 
we eliminate $\lambda_c^{(d)}$ through the unitarity of the CKM matrix -- 
instead of $\lambda_t^{(d)}$, as done in (\ref{Bdrhorho-ampl}) -- we obtain 
\begin{equation}\label{Bdrhorho-ampl2}
A(B_d^0\to[\rho^+\rho^-]_f)=\left(1-\frac{\lambda^2}{2}\right)\lambda^3
A\,R_b\,e^{i\gamma}\left[\tilde A_{\rm cc}^{(u)f}+\tilde A_{\rm pen}^{(uc)f}
\right]\left[1+r_f\,e^{i\sigma_f}e^{-i(\beta+\gamma)}\right],
\end{equation}
where
\begin{equation}\label{rf-def}
r_f\,e^{i\sigma_f}\equiv\frac{R_t}{(1-\lambda^2/2)R_b}
\left[\frac{\tilde A_{\rm pen}^{(tc)f}}{\tilde A_{\rm cc}^{(u)f}+
\tilde A_{\rm pen}^{(uc)f}}\right],
\end{equation}
with 
\begin{equation}
R_t\equiv\frac{1}{\lambda}\left|\frac{V_{td}}{V_{cb}}\right|={\cal O}(1). 
\end{equation}
Taking into account that we have $\phi_d=2\beta$ and 
$\beta+\gamma=180^\circ-\alpha$ within the Standard Model, we arrive at 
\begin{equation}
b_f=r_f,\quad \rho_f=\sigma_f,\quad\omega=\alpha,
\end{equation}
and at the ``effective'' mixing phase $\phi=\phi_d+2\gamma=-2\alpha$. 
Consequently, using the strategy presented in Section \ref{STRAT}, the 
$B_d\to\rho^+\rho^-$ angular distribution allows us to probe also the 
combination $\alpha=180^\circ-\beta-\gamma$ directly, i.e.\ to determine
$\alpha$ as a function of a CP-conserving strong phase difference 
$\Delta_{\tilde f,f}$ (see also (\ref{R-approx}) and (\ref{ID-approx})). 
Needless to note that the decay $B_d\to\rho^0\rho^0$ may also be 
interesting in this respect. Since the normalization factors ${\cal N}_f$ 
of the parametrization (\ref{Bdrhorho-ampl2}) are proportional to 
\begin{equation}
\tilde A_{\rm cc}^{(u)f}+\tilde A_{\rm pen}^{(uc)f},
\end{equation}
which is governed by ``colour-allowed tree-diagram-like'' topologies, 
it may well be that $\Delta_{\tilde f,f}\approx0$. This relation would allow 
us to extract $\alpha$, as well as the $B_d\to\rho^+\rho^-$ hadronic 
parameters, which include also another strong phase difference 
$\Delta_{f',f}$, providing an important cross check.

\boldmath
\subsection{The Decays $B_d\to K^{\ast0}\,\overline{K^{\ast0}}$ and
$B_s\to K^{\ast0}\,\overline{K^{\ast0}}$}\label{BdKK}
\unboldmath
The decays $B_{d}^0\to K^{\ast0}\,\overline{K^{\ast0}}$ and 
$B_{s}^0\to K^{\ast0}\,\overline{K^{\ast0}}$ are pure ``penguin'' modes,
originating from $\bar b\to\bar d s\bar s$ and $\bar b\to\bar s d\bar d$
quark-level transitions, respectively. They do not receive contributions 
from current--current operators at the ``tree'' level, and can be 
parametrized within the Standard Model in complete analogy to 
(\ref{Bd-ampl}) and (\ref{Bd-ampl2}). We have just to set the 
current--current amplitudes equal to zero in these expressions. 
The decays $B_{d}^0\to K^{\ast0}\,\overline{K^{\ast0}}$ and 
$B_{s}^0\to K^{\ast0}\,\overline{K^{\ast0}}$ are related to each other 
by interchanging all down and strange quarks, i.e.\ through the $U$-spin 
flavour symmetry of strong interactions, and the strategies to probe $\gamma$ 
and the $B^0_{d,s}$--$\overline{B^0_{d,s}}$ mixing phases are analogous to 
those discussed in Section~\ref{BPSI}. Since the $B_{d,s}\to K^{\ast0}\,
\overline{K^{\ast0}}$ decays are pure ``penguin'' modes, they represent a 
particularly sensitive probe for new physics.

An interesting alternative to parametrize the $B_{d}^0\to [K^{\ast0}\,
\overline{K^{\ast0}}]_f$ decay amplitudes within the Standard Model 
is given as follows:
\begin{equation}\label{BdKK-ampl}
A(B_d^0\to[K^{\ast0}\,\overline{K^{\ast0}}]_f)=\lambda^3A\,R_t\,
{\cal A}_{\rm pen}^{(tu)f}\,e^{-i\beta}\left[1-g_f\,e^{i\varphi_f}
e^{i\beta}\right],
\end{equation}
where
\begin{equation}
g_f\,e^{i\varphi_f}\equiv\frac{1}{R_t}
\frac{{\cal A}_{\rm pen}^{(cu)f}}{{\cal A}_{\rm pen}^{(tu)f}}
\end{equation}
may well be sizeable due to the presence of final-state-interaction 
effects \cite{RF3}. Consequently, we have 
\begin{equation}
b_f=g_f,\quad \rho_f=\varphi_f,\quad\omega=\beta.
\end{equation}
Because of the factor of $e^{-i\beta}$ in front of the square brackets 
on the right-hand side of (\ref{BdKK-ampl}), the ``effective'' mixing phase 
is given by $\phi=\phi_d-2\beta$. Consequently, the strategy presented in 
Section~\ref{STRAT} allows us to probe the CP-violating weak phase $\beta$
of the CKM element $V_{td}=|V_{td}|e^{-i\beta}$. Within the Standard Model, 
we have $\phi=\phi_d-2\beta=0$. However, this relation may well be affected 
by new physics, and represents a powerful test of the Standard-Model 
description of CP violation (for a recent discussion, see \cite{wolf-NP}). 
Therefore it would be very important to determine this combination 
of CKM phases experimentally. The observables of the 
$B_d\to K^{\ast0}[\to\pi^-K^+]\,\overline{K^{\ast0}}[\to\pi^+K^-]$
angular distribution may provide an important step towards this goal.

\boldmath
\section{Conclusions}\label{CONCL}
\unboldmath
The angular distributions of certain quasi-two-body modes $B_{d,s}\to 
X_1\,X_2$, where both $X_1$ and $X_2$ carry spin and continue to decay 
through CP-conserving interactions, provide valuable information about 
CKM phases and hadronic parameters. We have presented the general
formalism to accomplish this task, taking into account also penguin
contributions, and have illustrated it by having a closer look at a
few specific decay modes. In comparison with strategies using non-leptonic
$B_{d,s}$ decays into two pseudoscalar mesons, an important advantage of 
the angular distributions is that they provide much more information, 
thereby allowing various interesting cross checks, for instance, of certain 
flavour-symmetry relations. Moreover, they provide a very fertile testing 
ground for model calculations of the $B_{d,s}\to X_1\,X_2$ modes.

We have pointed out that the decay $B_d\to J/\psi\,\rho^0$ can be combined 
with $B_s\to J/\psi\,\phi$ to extract the $B^0_d$--$\overline{B^0_d}$ mixing 
phase $\phi_d=2\beta$ and -- if penguin effects in the former mode should 
be sizeable -- also the angle $\gamma$ of the unitarity triangle. As an 
interesting by-product, this strategy allows us to take into account also 
the penguin effects in the extraction of the $B^0_s$--$\overline{B^0_s}$ 
mixing phase from $B_s\to J/\psi\,\phi$. If penguin effects should be very 
small in $B_d\to J/\psi\,\rho^0$, $\phi_d$ could still be determined and it
would even be possible to resolve a twofold ambiguity, arising in the 
extraction of this CKM phase from $B_d\to J/\psi\,K_{\rm S}$. Other 
interesting applications, involving $B_d\to\rho \rho$ and 
$B_{s,d}\to K^{\ast}\overline{K^\ast}$ decays, were also noted. Within the 
Standard Model, these modes are expected to exhibit branching ratios at 
the $10^{-5}$ level; also the one for $B_d\to K^{\ast0}\overline{K^{\ast0}}$ 
may well be enhanced, from its ``short-distance'' expectation of 
${\cal O}(10^{-6})$ to this level, by final-state-interaction effects. 

Since the formalism presented in this paper is very general, it can of 
course be applied to many other decays. Detailed studies are required to 
explore which channels are most promising from an experimental point of 
view. Although the $B_d$ modes listed above may already be accessible at 
the asymmetric $e^+$--$\,e^-$ $B$-factories operating at the $\Upsilon(4S)$ 
resonance, which will start taking data very soon, the strategies presented 
in this paper appear to be particularly interesting for 
``second-generation'' experiments at hadron machines, such as LHCb or BTeV, 
where also the very powerful physics potential of the $B_s$ system can be 
exploited. 

\vspace{0.5cm}

\noindent
{\it Acknowledgement}

\vspace{0.1cm}

\noindent
I am very grateful to Jerome Charles for interesting discussions.

\end{document}